\documentstyle[epsf,twocolumn,aps,pre]{revtex}
\begin{document}
\draft
\twocolumn[\hsize\textwidth\columnwidth\hsize\csname @twocolumnfalse\endcsname
\title{A Neutral Polyampholyte in an ionic solution}  
\author{Alexandre Diehl, Marcia C. Barbosa and Yan Levin}
\address{Instituto de F\'{\i}sica, Universidade Federal do Rio Grande do Sul \\  
Caixa Postal 15051, CEP 91501-970 \\  
Porto Alegre, RS, Brazil}  

\maketitle

\begin{abstract}  
The behavior of a neutral polyampholyte 
($PA$) chain with $N$  monomers, in an ionic solution, is analyzed
in the framework of the  full Debye-H$\ddot u $ckel-Bjerrum-Flory $(DHBjF)$ theory. 
A $PA$ chain, that in addition to the neutral monomers, also contains an  equal number
of positively and negatively charged monomers, is dissolved  in an
ionic solution. For \underline{high} concentrations
of salt and at high temperatures, the $PA$ exists in 
an extended state. As the temperature is decreased,  the
electrostatic energy becomes more relevant and
at a $T=T_{\theta}$ the system collapses into a dilute globular state, 
or microelectrolyte. This state contains a concentration of salt higher than 
the surrounding medium. As the temperature is
decreased even further, association between the
monomers of the polymer and the ions of the salt
becomes relevant and there is a crossover
from this globular state to a low temperature
extended state. For \underline{low} densities of salt, the 
system is collapsed for almost all temperatures and
exhibits a first-order phase  transition to an extended state
 at an unphysical low temperature.
\end{abstract}
\vskip2pc]
\narrowtext

\section{\bf INTRODUCTION}

In recent years, there has been a lot of interest in the study of the physical properties of 
macromolecules. When dissolved in a good solvent, a neutral 
macromolecule, that is made of $N$ units (monomers), is characterized by a radius 
of gyration that scales as $R\sim N^{\nu} a$, with $\nu =3/5$. In a bad solvent, as the 
temperature is lowered, a polymer undergoes a phase transition from  an extended state, characterized by 
$\nu =3/5$, to a compact globular state, with $\nu =1/3$. At the transition point, $T=T_{\theta}$, 
the system assumes a coil configuration, with $\nu =1/2$. The transition resembles the tricritical 
behavior present in some magnetic materials\cite{DeG75} and can be described by the Flory-De Gennes 
theory\cite{DeG75,Fl53,DeG79,DeG72}.

In nature, however, charged polymers are prevalent. They can be 
divided into two classes, polyelectrolytes and polyampholytes ($PAs$).
The most common example of a polyelectrolyte is a DNA dissolved in water. Upon
dilution, the phosphate groups of the DNA  molecule 
lose their $H^+$ atoms and become ionized. All
the ionized groups of the polyion carry the charge of the same sign. The repulsion
between the like charges helps to stretch the molecule and
contributes to the overall persistence length of a polyion\cite{Fi79,Od79}. 

Proteins, on the other hand, are an example of polyampholytes. A PA molecule can 
have monomers of either positive or negative charge. Thus, two extremes are possible. In one 
extreme, we can have a neutral PA, in which the number of positive monomers is exactly 
equal to the number of negative monomers. In the other extreme, all the monomers can be of the 
same sign, in which case a PA will be equal to a polyelectrolyte. In this paper we shall 
confine our attention to the case of a neutral $PA$.

In contrast to a normal polymers, the understanding of $PA$s is rather poor.  
This is due to the difficulty of properly accounting for the long range Coulomb
interaction between the monomers of a $PA$. The long range electrostatic interaction is qualitatively
different from the short-range excluded volume repulsion, that drives the coil-globule transition
in an usual uncharged polymer. Consequently, our understanding of the full 
conformational structure of the $PA$ is rather incomplete.

One of the first attempts to study a neutral $PA$ was done by 
Edwards {\it et al.}\cite{Edwa80}. They argued, on the basis of the Debye-H$\ddot u$ckel ($DH$) 
limiting law, that when the polymer chain is in a nonionic solvent, the competition between the net 
attractive electrostatic energy of the $PA$ and the reduction in its entropy, due to the confinement 
in a sphere smaller than its natural radius, leads to a transition from an extended coil to a 
``dilute globular state'' (microelectrolyte). The microelectrolyte is characterized by a density 
smaller than that of the collapsed state of a chain made of neutral monomers. The microelectrolyte 
picture of a $PA$ was further extended and elaborated  in the work of Higgs and Joanny\cite{Hi91}, who 
have concluded that a sufficiently long $PA$ will always be collapsed.

In a recents papers, we have investigated\cite{Le95b,Barb95} the possibility of a different type of 
coil-globule transition in a neutral non-alternating $PA$. In the absence of salt, 
the behavior of a neutral $PA$, with $N$ charged monomers, was analyzed in the framework of the full 
$DH$ theory\cite{DH23}, augmented by Bjerrum's idea of ion association\cite{Bj26} and Flory's 
affine network theory\cite{Fl76} of rubber elasticity. In this approach, the allowance was made for the 
fact that a strong electrostatic attraction between the unlike monomers makes it favorable 
for them to associate forming intermolecular bridges. The $PA$ molecule then resembles an 
affine network with the crosslinks of functionality $4$. It was found that at high temperatures 
the crosslinks are quite rare, and the picture of a $PA$ as existing in a dilute globular 
state remains unchanged, with the radius of gyration scaling as $R\sim aN^\nu $ and $\nu=1/3$ 
(microelectrolyte), as predicted by Edwards {\it et al.}\cite{Edwa80,Hi91}. As the temperature is 
decreased, the ionic association becomes energetically
favorable. At very low temperatures, the fraction of unassociated charged 
monomers approaches zero and the $PA$ molecule resembles a microgel, with $\nu =2/5$
(compare with the Flory exponent  $\nu_{F}=3/5$). The two states 
are separated by a first order phase transition. It is interesting to note that unlike the case of an usual 
neutral polymer, the extended state of $PA$ $(\nu =2/5)$ occurs at low temperatures.

In order to analyze the effects of addition of salt to the polymeric solution, Higgs and Joanny\cite{Hi91} 
extended Edwards {\it et al.}\cite{Edwa80} method, including the electrostatic interaction between the
charged monomers of the chain and the free ions of the salt. They 
considered the case where the density of charges on the $PA$ is
of the same order of magnitude as the density of added salt. 
They have concluded that the only effect of added
salt was to renormalize the second virial coefficient, adding an attractive term to it. At high temperatures and in
a good solvent, the excluded volume term dominates
the attractive $DH$  contribution and  the chain is extended. As the
temperature is decreased, 
the electrostatic contribution  balances the excluded volume contribution
and the polymer collapses into a dilute globular state.

Higgs and Joanny's theory (HJ), however, is based on the assumption that 
the  density of the polymer is close to the density of salt, what is actually a good 
approximation  only when the densities are quite high.
Furthermore, since HJ work in the framework of the linearized $DH$ theory, their 
conclusions can only be trusted at  high temperatures and cannot be extended to the low temperatures, 
where they find an unphysical divergence of the concentration of ions near the $PA$. 

In order to further explore these issues, we analyze this problem in the framework of the full 
Debye-H$\ddot u$ckel Bjerrum-Flory $(DHBjF)$. The possible association between the counterions 
and the monomers of the chain is explicitly taken into account.
In a good solvent and for density of salt $\rho_{0} > \bar \rho$ ($\bar \rho$ is the density of 
polymer in extended state at high temperatures), corresponding to the region 
where HJ found the transition, the system has the following behavior. For high temperatures,
the repulsion due to the  excluded volume dominates over the 
$DH$ electrostatic attraction and  the polymer is 
extended. The counterion association is insignificant in this region. As the temperature is decreased, 
the electrostatic energy prevails and at $T=T_{\theta}$ the system 
collapses into the dilute globular state, in agreement with HJ predictions. We obtain explicitly the
values of  the transition temperature for different densities of salt and for different volume fractions of 
charged monomers. As the temperature is lowered even further, the counterion-monomer association becomes energetically favorable. 
The number of ions of salt attached to the chain increases. As $T \rightarrow 0$, the 
fraction of unpaired monomers approaches zero and the $PA$ molecule becomes
extended like an usual neutral polymer. Within our approach we find that the instability observed by HJ is hidden 
inside the metastable region of the salt solution. Furthermore, beyond the HJ results, we are able to investigate the 
properties of a polymer-salt solution when the density of salt is quite small. We find that for $\rho_{0} < \bar \rho$, the
fluctuation induced attraction between the monomers is strong and the polymer exists in a dilute collapsed state for 
{\it almost} all temperatures. At very low temperatures and low densities, the
system undergoes a first-order transition into an extended state. 

The paper is organized as follows. In Sec.\ref{model} our model for $PA$+salt is presented and the 
Helmoltz free energy of the composed system is constructed. A full thermodynamic analysis is formulated 
in Sec.\ref{therm}, in particular, the equations that determines the equilibrium configurations of the system. In Sec.\ref{results} a detailed discussion of our results is presented.

\section{\bf DESCRIPTION OF THE MODEL}
\label{model}
 
\bigskip
\subsection{\bf Introduction} 
\bigskip

We consider a polyampholyte ($PA$) chain consisting of $N$
constitutional units (hereafter referred to as monomers) of
three different types: $N_N$ neutral monomers, $(N-N_N)/2$
positive monomers and $(N-N_N)/2$ negative monomers, randomly 
distributed along the chain, each spherical in shape with
a diameter $a$. The polymer chain  occupies an effective sphere of
volume $V_{i}=4\pi R^3/3$, where $R$, the radius of gyration,
measures the end-to-end extension of the polymer. The
reduced density of monomers, $\rho^{\ast}=\rho a^3$, inside the globule is
$\rho^{\ast} =3Na^{3}/(4\pi R^3)$.

The solvent that occupies the total volume $V$ of the system 
is represented by an uniform medium of a dielectric constant $\epsilon$ in which
an amount of simple 1:1 salt has been added.
The positive and negative ions of salt dissolved in the solvent have a fixed densities 
$\rho_0=\rho_0^+ + \rho_0^- = 2\rho_0^+ =2\rho_0^-$. The basic approximations of our model are:

1. In the limit of large dilution we can neglect the interactions between different $PAs$ and 
concentrate our attention on one chain.

2. The most relevant interaction between the monomers of the chain and the free ions of salt is
of electrostatic nature. For simplicity,  the excluded volume interactions between the
$PA$ and the salt will be 
neglected. 

3. The short range interactions between the polymer and nonionic solvent are included 
in the effective value of the second virial coefficient. 

\begin{figure}[h]
\begin{center}
\epsfxsize=6.cm
\leavevmode\epsfbox{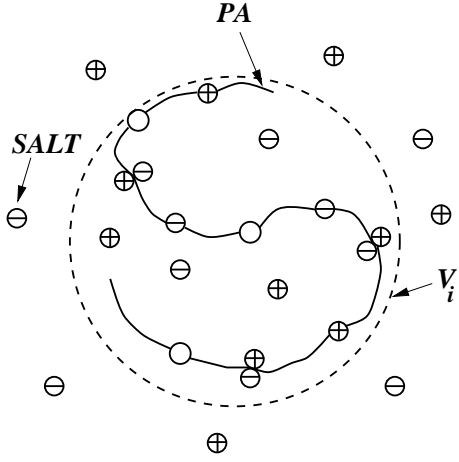}
\end{center}
\caption{Schematic drawings of a $PA$ chain with $N$ neutral monomers $\bigcirc$, 
$(N-N_{N})/2$ positive monomers $\oplus$, and $(N-N_{N})/2$ monomers $\ominus$, immersed in an 
ionic solution. The salt ions can associate with the positive and negative monomers 
of the chain, forming dipolar pairs. The $PA$ is confined to a spherical region of radius $R$.}
\label{globule}
\end{figure}

The strong electrostatic interactions make it favorable for the ions of salt to associate
with the $PA$. After the thermal equilibrium is achieved, the chain is composed
of three different structures: neutral monomers, free charged monomers and monomer that have 
an associated ion of salt attached to them (see Fig.\ref{globule}). We shall refer to the latter as a 
``dipolar pair''. The total number of monomers can then be written as
\begin{equation}
\label{Ntot}
N=N_N + N_p^+ + N_p^- + N_d\;,   
\end{equation}
where $N_p^{+}$ and $N_p^{-}$ are the number of unassociated positive and negative  
monomers and $N_d$ is the total number of dipoles. Defining $f$ as the
number fraction of charged monomers in the $PA$, and $x$ as the
fraction of charged monomers that form dipoles, we can write 
\begin{eqnarray}  
\label{Nmon} 
N_N &\equiv&(1-f)N\;,\qquad \qquad \qquad
\mbox{neutral,}\nonumber\\
N_p^{+}&\equiv&\frac{f}{2} (1-x)N=N_p^{-}\;,\qquad
\mbox{charged,}\nonumber\\ N_d&\equiv&fxN\;,\qquad \qquad \qquad
\qquad \mbox{dipoles.} 
\end{eqnarray}

It is then possible to express the densities of neutrals, charges and 
dipolar pairs, in terms of the total density of the chain, $\rho\equiv N/V_{i}$, by the expressions 
\begin{eqnarray}  
\label{Ndens} 
\rho_N &\equiv&(1-f)\rho\;,\nonumber\\ 
\rho^{+}&\equiv&\frac{f}{2} (1-x)\rho=\rho^{-}\;,\nonumber\\ 
\rho_d &\equiv& fx\rho =\rho_d^+ + \rho_d^- \;, 
\end{eqnarray}
where $\rho_d^+$ and $\rho_d^-$ are the densities of dipoles containing 
positive and negative monomers, respectively. In principle, ions of
salt are different from the monomers of the $PA$ and the dipoles formed of positive and
negative monomers have to be distinguished. Since, however, the $PA$ is neutral, in equilibrium
$\rho_d^+ =\rho_d^-$.

The salt particles are free to move along the whole volume of
the composed system. In the absence of the $PA$, the positive and
negative charges are uniformly distributed with a constant density $\rho_0$. In the presence of $PA$, 
due to electrostatic fluctuations, the density of salt near the the polymer is \underline{not} equal 
to $\rho _0$. Therefore, in order to treat the problem in a simple way, we shall divide the 
total system of volume $V$ into two subregions: ``inside'' and ``outside'' the volume $V_{i}$. The 
density of salt particles outside is uniform and equal to $\rho _0=N_{0}/V_0$, where $V=V_i+V_0$. 
The density of the salt inside  the volume $V_{i}$ is $\rho_s =\rho_s^{+}+\rho _s^{-}$. Since there is 
an equal number  of positive and negative free ions, $\rho_s^{+}=\rho_s^{-}=\rho_s /2 \equiv N_{s}/V_{i}$.

\bigskip
\subsection{\bf  Helmoltz free energy}
\bigskip

The Helmoltz free energy for the whole system is the sum of two terms, $F=F_i+F_0$. The ``inside'' 
free energy $F_i$, associated with the subregion where the polymer and the salt coexist, and the ``outside'' 
contribution $F_0$, for the regions where only salt is present. The free energy in the outside region is 
composed of:

\noindent . an energy due to short range repulsion;

\noindent . an electrostatic energy due to the interaction between ions;

\noindent . and the kinetic-entropic contribution. 

\noindent For the inside region, the Helmoltz free energy $F_i$, contains:
 
\noindent . an elastic contribution due to the elongation and contraction of the chain;

\noindent . an energy due to the monomer-monomer interaction;

\noindent . an energy due to the ion-ion short range repulsion;

\noindent . an electrostatic energy of monomer-monomer, salt-salt and monomer-salt interactions.

\noindent . an entropic energy due to the mixing of  the different types of particles present
inside the region $V_i$. 

\bigskip
\noindent {\bf a. Elastic free energy} 
\bigskip

According to the Flory-De Gennes\cite{Fl53,Fl76,Al76} theory, the elastic free energy of the chain is\cite{DeG76}  
\begin{equation} 
\label{Fel}
\beta F_{el}=\frac{3}{2}\left( \alpha^2 -1\right) -3\ln \alpha\;, 
\end{equation}
where $\alpha = R/R_0$ is the expansion factor of the chain, measured relative to the 
non-strained Gaussian state with radius $R_0\approx a\sqrt{N}$.  

\bigskip
\noindent {\bf b. Free energy of excluded volume}
\bigskip

Following the usual Flory-De Gennes\cite{DeG75,Fl53,DeG79,DeG72} theory, the short range 
interactions between the particles of the chain, at low densities, are approximated by a 
virial expansion
\begin{equation} 
\label{Fhcp}
\beta F_{HC}^p=\frac{N}{2}W_1 \rho + \frac{N}{2}W_2 \rho^2\;,  
\end{equation}
where $W_1$ and $W_2$ are second and the third virial coefficients.

The excluded volume interaction for the salt is approximated by the free-volume form \cite{YanA,Le93}
\begin{equation} 
\label{FhcI}
\beta F_{HC}^s= - V_i\rho_s \ln \left( 1 - B\rho_s \right)\;,  
\end{equation}
where $B$ is chosen to yield appropriate maximal packing densities or to match
high-T second virial coefficient. Since we are assuming that the relevant interaction between the 
monomers and the salt particles is electrostatic, we do not consider the excluded volume repulsion 
between the free ions inside $V_i$ and the monomers. Outside $V_i$, the excluded volume contribution to the
Helmoltz free energy is   
\begin{equation} 
\label{Fhc0}
\beta F_{HC}^0= - V_0\rho_0 \ln \left( 1 - B\rho_0 \right)\;,
\end{equation}
and, for small densities $\rho_0$, it can be neglected.

\bigskip
\noindent {\bf c. Electrostatic term} 
\bigskip

The electrostatic interaction between the charges is calculated in
the framework of the Debye-H$\ddot u$ckel ($DH$) theory. The standard argument leads to\cite{DH23,YanA,Le93,Le94} 

\begin{equation} 
\label{FDH}
\beta F_{DH}=-\frac{N_1}{T^{\ast}(\kappa a)^2} 
\biggl[\ln(\kappa a+1) + \frac{(\kappa a)^2}{2} - \kappa
a\biggr]\;.
\end{equation}
Inside $V_i$, the inverse Debye screening length $\kappa=1/\xi_{D}$, is
determined by all free unassociated charged species, i.e., both
monomers and salt ions, so that $(\kappa a)^2 =4\pi \rho_1^{\ast}/T^{\ast}$, where 
$\rho_1^{\ast}=a^3 N_1 /V_i = \rho_s^{\ast} + f(1-x)\rho^{\ast}$ and $T^{\ast}=k_{B}T\epsilon a/q^2$. 
Outside the volume $V_i$, the Debye length is determined by the concentration of salt, $(\kappa a)^2 =(\kappa_0 a)^2= 4\pi \rho_0^{\ast}/T^{\ast}$. 

The $DHBj$ theory has proven to be successful in giving the correct qualitative picture 
of a simple electrolyte\cite{YanA,Le93,Le94}. Thus, while inclusion of higher order contributions, 
such as dipole-ion and dipole-dipole interactions, improves the quantitative agreement of the theory 
with experiments\cite{Pi92} and Monte Carlo simulations\cite{Panan92}, most of  the important physics is 
already captured at the $DHBj$ level\cite{Le93}. As a leading order approximation we shall, therefore, 
treat the dipolar clusters as ideal noninteracting specie.

\bigskip
\noindent {\bf d. Free energy of mixing}  
\bigskip

Inside the volume $V_i$, the system is composed of a mixture of neutral, positive and negative monomers, as 
well as, positive and negative ions of salt, and dipoles made of monomers and associated counterion. If the 
bonds connecting the monomers were cut, we would have a ``gas'' made of these
species. The ideal gas free energy of the mixture is
\begin{equation} 
\label{Fid}
\beta F_{id}=\sum_j\biggl[N_j\ln \frac{\rho _j\Lambda
_j^{3k_j}}{\xi _j}-N_j\biggr]\;, 
\end{equation}
where $\Lambda _j(T)$ is the mean thermal wavelength, $\Lambda_j(T)=h/ \sqrt{2\pi \bar m_jk_BT}$, 
$\bar m_j$ is the geometric mean mass for a cluster of $k_j$ units and $\xi _j(T)$
is the internal partition function. The index $j$ specifies all the possible species: positive, 
negative, and neutral monomers, dipoles, and positive and negative free ions. In our case, $k_N =k_-=k_+=k_s=1$ 
and $k_2=2$. To simplify the notation, we shall take the
masses of both monomers and ions to be equal to $m$; therefore, $\Lambda_j(T)=\Lambda$. Since all the monomers 
are hard spheres, the molecular partition is a constant, $\xi_{N}=\xi_{1}=\xi_s =\xi =1$. 
For dipoles, however, the internal partition function is\cite{Bj26} 
\begin{equation} 
\label{KT}
\xi_{2}(T^{\ast})=4\pi \int_{a}^{c} e^{a/T^{\ast}}r^2 dr \equiv K(T^{\ast})\;,  
\end{equation}
where $K(T)$ defines an association constant for the formation of
bound dipolar pairs. For the cutoff $c$, we use the Bjerrum's
choice, $a/2T^{\ast}$, which corresponds to the inflection
point of the integral as a function of $c$. The Bjerrum equilibrium constant is 
\begin{eqnarray}
\label{KBj}
K(T^{\ast})&=&\frac{2\pi a^{3}}{3(T^{\ast})^3}\biggl[Ei\left(\frac{1}
{T^{\ast}}\right)-Ei\left(2\right)+e^2\biggr]\nonumber\\
&&-\frac{2\pi a^3}{3}e^{1/T^{\ast}}\biggl[2+\frac{1}{T^{\ast}}+\frac{1}{(T^{\ast})^2}\biggr]\;. 
\end{eqnarray}
As was shown by Ebeling\cite{Ebel68} the low temperature asymptotic value of this constant is in 
\underline{exact} agreement with the cluster expansion\cite{KEbel}. 

While the ions of the salt are free to move independently of
each other, the monomers of PA are constrained by the bonds
that maintain the integrity of the chain. 
The equation (\ref{Fid}), therefore, overestimate the entropic free energy by the amount equivalent 
to the the free energy of the ideal gas composed of $N$ particles. The free energy of mixing is therefore, 
\begin{eqnarray} 
\label{Fmix}
\beta F_{MIX}&=&\sum_{j}\biggl[N_{j}\ln \frac{\rho_{j}
\Lambda_{j}^{3k_j}}{\xi_j} -N_{j}\biggr]-\biggl[N\ln \frac{\rho
\Lambda^3}{\xi} -N\biggr] \nonumber \\
&=&V_i\rho_{N}\ln \Lambda^3 \rho_{N} + V_i\rho^+ \ln \Lambda^3 \rho^+ + 
V_i\rho^-\ln \Lambda^3 \rho^-\nonumber\\
&& + V_i\rho_{d}^{+}\ln \Lambda^6 \frac{\rho_{d}^+}{K(T)}
+V_i\rho_{d}^{-}\ln \Lambda^6 \frac{\rho_{d}^{-}}{K(T)}\nonumber\\ 
&&+ V_i\rho_{s}^{+}\ln \Lambda^3 \rho_{s}^{+} + V_i\rho_{s}^{-}\ln \Lambda^3 \rho_{s}^{-}\nonumber\\
&&-V_i\rho \ln \Lambda^3 \rho - V_i\rho_s^+ - V_i\rho_s^-\;.
\end{eqnarray}

Outside, in $V_0$, the entropic contribution to the free energy is given by an ideal gas free energy 
\begin{equation} 
\label{MIX0}
\beta F_{id}^{0}= V_0\rho_{0}^{+}\ln \Lambda^3 \rho_{0}^{+} + V_0\rho_{0}^{-}\ln \Lambda^3
\rho_{0}^{-} - V_0\rho_0^+ - V_0\rho_0^-\;.  
\end{equation}

\bigskip
\section{\bf THERMODYNAMICS OF THE SYSTEM}
\label{therm}
\bigskip

The equilibrium configuration of the $PA+$salt system is determined by the minimization of the Gibbs free energy.
Since the total system is composed of two subregions, the total
Gibbs free energy can be written as  $G\equiv G_i+G_0$, where $G_i$
is the Gibbs free energy of the subsystem formed by the
``inside'' region and $G_0$ is the Gibbs free energy for ``outside'' region, $V_0$. The minimization of 
the total free energy implies that $\delta G\equiv\delta G_i+\delta G_0=0$ and, consequently,
\begin{eqnarray} 
\label{miniGT} 
0&=&\frac{\partial G_i}{\partial V_i}dV_i 
+ \frac{\partial G_0}{\partial V_0}dV_0 
+ \frac{\partial G_i}{\partial N_p^+}dN_p^+ 
+ \frac{\partial G_i}{\partial N_p^-}dN_p^-\nonumber\\
&&+ \frac{\partial G_i}{\partial N_d^+}dN_d^+
+ \frac{\partial G_i}{\partial N_d^-}dN_d^- 
+ \frac{\partial G_i}{\partial N_s^+}dN_s^+\nonumber\\
&&+ \frac{\partial G_i}{\partial N_s^-}dN_s^-
+ \frac{\partial G_0}{\partial N_0^+}dN_0^+ 
+ \frac{\partial G_0}{\partial N_0^-}dN_0^-\;.
\end{eqnarray}

The total volume of the system is fixed, $V =V_i + V_0$, which
leads to the condition $\delta V_i=-\delta V_0$ and, consequently, to the equality of the osmotic 
pressures  $P\equiv -\partial G/\partial V$ inside and outside the globule, 
\begin{equation} 
\label{pressure}
P_i \equiv P_p + P_s =P_0\; .
\end{equation}
In the above expression, our first condition for equilibrium,
$P_p$,$P_s$ and $P_0$ are the osmotic pressures of the polymer,
salt inside  the volume $V_i$,  and salt in $V_0$, respectively. 

Now, since the total number of ions of salt, $N_s^- + N_s^+ + N_d^+ + N_d^- + N_0^- + N_0^+$, 
and the total number of monomers that contain charges, $N_p^- + N_p^+ + N_d^- + N_d^+$ 
(inside $V_i$), are constants, we have $\delta N_p=-\delta N_d$ 
(where $N_d=N_d^++N_d^-$  and $N_p=N_p^-+N_p^+$) and $\delta N_0=-\delta N_d-\delta N_s$ 
(where $N_s=N_s^-+N_s^+$ and $N_0=N_0^-+N_0^+$). 
Furthermore, since we are considering a neutral $PA$ and a neutral salt solution, the 
chemical potentials, $\mu_s^{\pm}\equiv\partial G/\partial N_s^{\pm}$ and
$\mu_p^{\pm}\equiv\partial G/\partial N_p^{\pm}$ , for negative
and positive species are equal, $\mu_s^-=\mu_s^+=\mu_s$ and $\mu_p^-=\mu_p^+=\mu_p$. Therefore, Eq.(\ref{miniGT}) becomes
\begin{equation} 
\label{miniGT1}
0=- \mu_p dN_d + \mu_d dN_d + \mu_s dN_s - \mu_0 ( dN_s+dN_d)=0\;,
\end{equation}
where $\mu_p$ and $\mu_d$ are the chemical potentials
of the charged monomers and the dipoles along the chain and $\mu_s$ and $\mu_0$ are the chemical 
potentials of free salt inside and outside the volume $V_i$, respectively. 
Since the number of dipole pairs and free ions are independent variables, the Eq.(\ref{miniGT1})
leads to two additional equilibrium conditions: the chemical association,
\begin{equation} 
\label{reaction}
\mu_p + \mu_s = \mu_d\;, 
\end{equation}
and the condition for free exchange of salt between the ``inside'' and the ``outside'' regions,
\begin{equation} 
\label{action}
\mu_0 = \mu_s\;. 
\end{equation}

Instead of working with a full system composed of both 
the ``inside'' and the ``outside'' regions, we can confine our 
attention strictly to the ``inside'' region, at the expanse of introducing some Lagrange multipliers. This leads to an \underline{effective} Gibbs free energy 
\begin{equation} 
\label{gibbs}
\beta {\cal G}=\beta F+\beta P_0V-\beta \mu _0N_s-\beta \mu_0N_d\;, 
\end{equation}
where $F$ is the Helmoltz free energy of the ``inside'' region,
\begin{equation} 
\label{Helm}
\beta F=\beta F_{el}+\beta F_{HC}^p+\beta F_{HC}^s+\beta F_{DH}+
\beta F_{MIX}\;,
\end{equation}
and the pressure $P_0$ and the chemical potential $\mu_0$ play the role 
of Lagrange multipliers. All the thermodynamic information about the
``inside'' region is now contained in $\cal{G}$. The equilibrium state of the  $PA+salt$ subsystem
is determined by the minimization of $\cal{G}$, $\delta {\cal G}=0$
\begin{eqnarray} 
\label{MiniGT2}
&&\left(\frac{\partial \beta F}{\partial V} +\beta P_0 \right) dV
+\left(\frac{\partial \beta F}{\partial N_s}-\beta \mu_0 \right) dN_s\nonumber\\
&&\qquad+\left(\frac{\partial \beta F}{\partial N_d} -\beta \mu_0 \right) dN_d
+\frac{\partial \beta F}{\partial N_p} dN_p=0\;.
\end{eqnarray}
Making once again use of the conservation of the number of charged
monomers and of the number of particles of salt, it is not hard to show that this reduce to three 
equilibrium equations (\ref{pressure}), (\ref{reaction}) and (\ref{action}), respectively. 

To find the quilibrium state of the $PA+$salt system, these equations must be solved simultaneously. 
Explicitly, the conditions for equality of pressures and for free exchange of salt between the 
``inside'' and the ``outside'' regions are respectively
\begin{eqnarray} 
\label{press} 
\rho_s &+& \frac{B\rho_s^2}{1-B\rho_s} +\frac{1}{8\pi a^3}\biggl[2\ln
(z+1)-\frac{z(z+2)}{z+1}\biggr]\nonumber\\
&&-\frac{3(\alpha^2 -1)}{4\pi a^3 \alpha^3
N^{3/2}} +\frac{W_1}{2}\rho^2 =\rho_0 + \frac{B\rho_0^2}{1-B\rho_0}\nonumber\\
&&+\frac{1}{8\pi a^3}\biggl[2\ln (z_0
+1) -\frac{z_0(z_0 +2)} {z_0+1}\biggr]\;, 
\end{eqnarray}
where $z\equiv \kappa a=\sqrt{4\pi \rho_1^{*}/T^{*}}$ and
$z_0\equiv \kappa_0a=\sqrt{4\pi \rho_0^{*}/T^{*}}$ and,
\begin{eqnarray} 
\label{chem}
\frac{(1-B\rho_s )\rho_0}{(1-B\rho_0 )\rho_s}&=&\exp \biggl[
-\frac{1}{2T^{\ast}}\biggl[\frac{z}{z+1}- \frac{z_0}{z_0 +1}\biggr]
+\frac{B\rho_s}{1-B\rho_s}\nonumber\\
&&\qquad \quad -\frac{B\rho_0}{1-B\rho_0}\biggr]\;.
\end{eqnarray}

The condition of chemical equilibrium between the dipoles and the monopoles reduces to,
\begin{equation} 
\label{reac}
\frac{2x}{(1-x)\rho_s} = \frac{K(T^{\ast} )}{1-B\rho_s} \exp
\biggl[-\frac{1}{T^{\ast}}\frac{z}{z+1} +
\frac{B\rho_s}{1-B\rho_s}\biggr]\;.  
\end{equation}

The equilibrium configuration of the $PA$, for a fixed
$T$ and $\rho_0$, is given by the solutions of the coupled
Eqs.(\ref{press}), (\ref{chem}) and (\ref{reac}).

\bigskip
\section{\bf RESULTS AND DISCUSSION}
\label{results}
\bigskip

\begin{figure}[h]
\vspace*{6cm}
\includegraphics{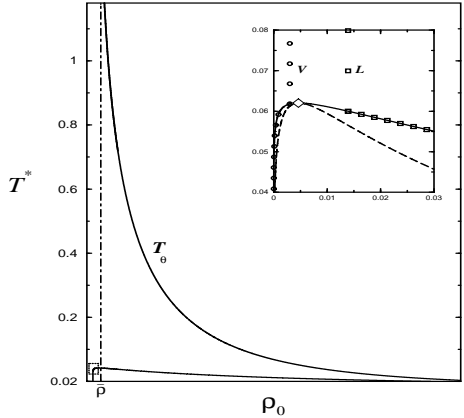}
\caption{The coexistence curve for the simple 1:1 salt, for the restricted primitive model (RPM) according 
to Debye-H$\ddot u $ckel (DH) theory, with hard-core repulsive contributions. The dotted box is 
represented in the inset. The critical point (open diamond) is $T_c^{\ast} \approx 0.0624$ and $\rho_c^{\ast} \approx 0.00462$, which implies two paths for the density of salt outside $V_i$: $(V)$ on the vapor side of 
coexistence curve (open circles) and $(L)$ on the liquid side of coexistence curve (open squares). The dashed 
line represents the HJ line of instability, and the dot-dashed line is the minimum density $\bar \rho$ of the polymer. The solid line represents the $\theta$ temperature $T_{\theta}$.}
\label{CCurve}
\end{figure}

For a fixed temperature and the external density $\rho_0$, these
equations can now be solved to
yield the values of $\alpha$, $x$ and $\rho_s$.
However, as the temperature is lowered, below $T_c^{\ast}$ (see Appendix 
\ref{CC} for details), the salt solution phase separates into
a high (liquid) and low (vapor) density phases. A $PA$  then finds itself 
either in the high or low density phase of the salt solution 
(see Fig.\ref{CCurve}). We shall assume that the $PA$ will be in the high 
density phase if the original density $\rho_0^{\ast}>\rho_c^{\ast}$, where 
$\rho_c^{\ast}$ is the critical point density for the pure salt solution\cite{YanA,Le93},
and in the low density phase $\rho_0^{\ast}<\rho_c^{\ast}$.
We chose a path such that, as the temperature is lowered, the density of the salt remains constant and
equal to $\rho_0$ up to $T_{\sigma}^{\ast}$ (the temperature at which 
$\rho_0=\rho_{\sigma}(T_{\sigma}^{\ast})$, where $\rho_{\sigma}$ is the coexistence curve 
boundary). Below $T_{\sigma}^{\ast}$, $\rho_0=\rho_{\sigma}(T^{\ast})$.

In order to obtain the phase-diagram of the system, we assumed
$W_1 = 4\pi a^3 /3$, the value corresponding to the second
virial coefficient for the gas of hard spheres, $N=100$ and $f=1$.
We chose two different paths: one on the vapor side of the coexistence curve, represented by the path $V$ in 
Fig.\ref{CCurve} and another on the liquid side of the coexistence curve, 
represented by the path $L$.
 
\begin{figure}[h]
\begin{center}
\epsfxsize=5.5cm
\leavevmode\epsfbox{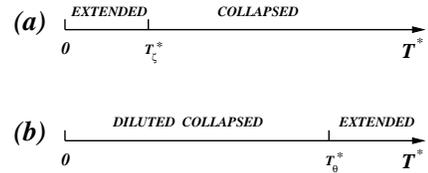}
\end{center}
\caption{Schematic phase-diagram for: (a) $\rho_0$ on the vapor side and 
(b) $\rho_0$ on the liquid side of the coexistence curve. The temperature 
$T_{\zeta}^{\ast}$ is very low.}
\label{Higgs}
\end{figure}

\begin{figure}[h]
\vspace*{12cm}
\includegraphics{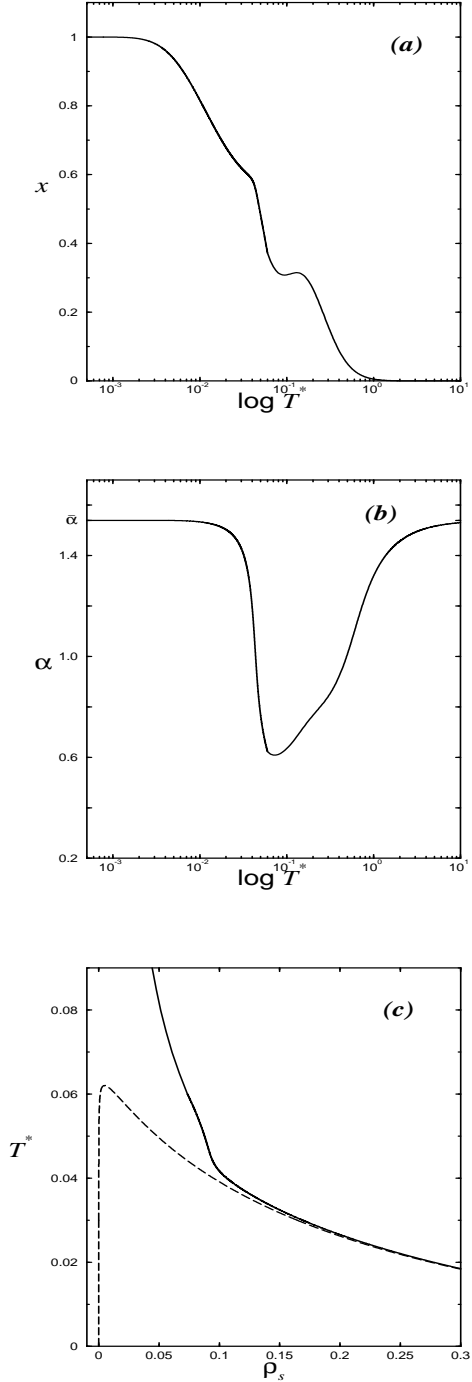}
\caption{(a) Fraction of dipoles $x$, (b) expansion factor $\alpha$
and (c) density of salt inside of spherical region, $\rho_s$, for
a quench on the liquid side of the coexistence curve, is shown against the temperature, for $N=100$, $f=1$ and 
$W_1 =4\pi a^3 /3$.} 
\label{LSCC1}
\end{figure}

First, the high temperature density of the salt outside the volume $V_i$
was fixed at $\rho_0 =0.0139 > \bar \rho$ (corresponding to $T_{\sigma}^{\ast}=0.06$), which 
implies a quench on the liquid side of the coexistence curve as shown in Fig.\ref{CCurve}. Here, $\bar \rho$ is the density of the polymer in the extended state, associated with $\alpha =\bar \alpha$, where $\alpha$ is the solution of Eq.(\ref{scale2}), with $W_1 >0$ (see Fig.\ref{LSCC1}(b)). Solving Eqs.(\ref{press}), (\ref{chem}) and (\ref{reac}), the following    
phase-diagram arises (see Fig.\ref{Higgs} and Fig.\ref{LSCC1}).
At high temperatures, the excluded volume repulsion is the dominant interaction and
the polymer is extended (scaling as $\nu=3/5$). As the temperature is decreased, the fluctuation induced 
electrostatic attraction increases and, at a temperature $T_{\theta}^{\ast}$ the polymer collapses into 
a dilute globular state, the microelectrolyte (see Appendix \ref{Temperature} for details). As the temperature is decreased 
even further, association between the monomer and the counterions begins to takes place. The values of 
$x$ and $\alpha$ smoothly increases and the polymer expands. Meanwhile, the density of the salt inside the 
globule approaches from above the value of the density 
outside the volume $V_i$, $\rho_{\sigma}^l(T)$. In the Appendix \ref{Temperature} this collapse 
transition is analyzed in detail and the transition temperature 
$T_{\theta}^{\ast}$ is calculated in the framework of the full $DH$ theory for different 
values of salt density and fraction of charged monomers. The unphysical divergence of the density of salt in 
the vicinity of a $PA$, encountered by HJ is no longer present, being hidden inside the coexistence region, as shown in Fig.\ref{CCurve}. 

Next, the high temperature density of salt outside $V_i$ was fixed at
$\rho_0^{\ast} = 0.003<\rho_{c}^{\ast}$ (at $T_{\sigma}^{\ast}=0.0618$), 
which implies a path on the vapor side of the coexistence curve. The solution of 
Eqs.(\ref{press}), (\ref{chem}) and (\ref{reac}) leads to the phase-diagram for 
this region, shown in Fig.\ref{Higgs} and Fig.\ref{VSCC1}. Even for high temperatures, the electrostatic energy is 
quite large and, consequently, the polymer is found in the dilute globular 
state for any physical temperature. Therefore, no high temperature collapse transition 
is found, when the density of salt is $\rho_0^{\ast} < \bar \rho$. As the temperature is lowered, 
however, the ionic association becomes
energetically favorable and the free ions of the salt
to condense onto the charged monomers of the $PA$. Indeed, in this region 
$x$, $\rho_s$ and $\alpha$ are not a single valued function. Besides the 
dilute globular state (designed with $d$ in Fig.\ref{VSCC1}), there are two 
other branches: one metastable (designed with $m$ in Fig.\ref{VSCC1}) and 
another stable branch (assigned by $l$ in Fig.\ref{VSCC1}). In
the presence of a good solvent, the polymer in this last stable branch is 
actually extended. At $T^{\ast}\approx T_{\zeta}^{\ast}$ (see Fig.\ref{Higgs}), 
there is a first-order phase transition between the dilute globular state 
and the low temperature extended state. This transition actually happens at 
such a low temperature (see Appendix \ref{Energy} for details), that it is even 
difficult to give a good estimate numerically. 

\begin{figure}[h]
\includegraphics{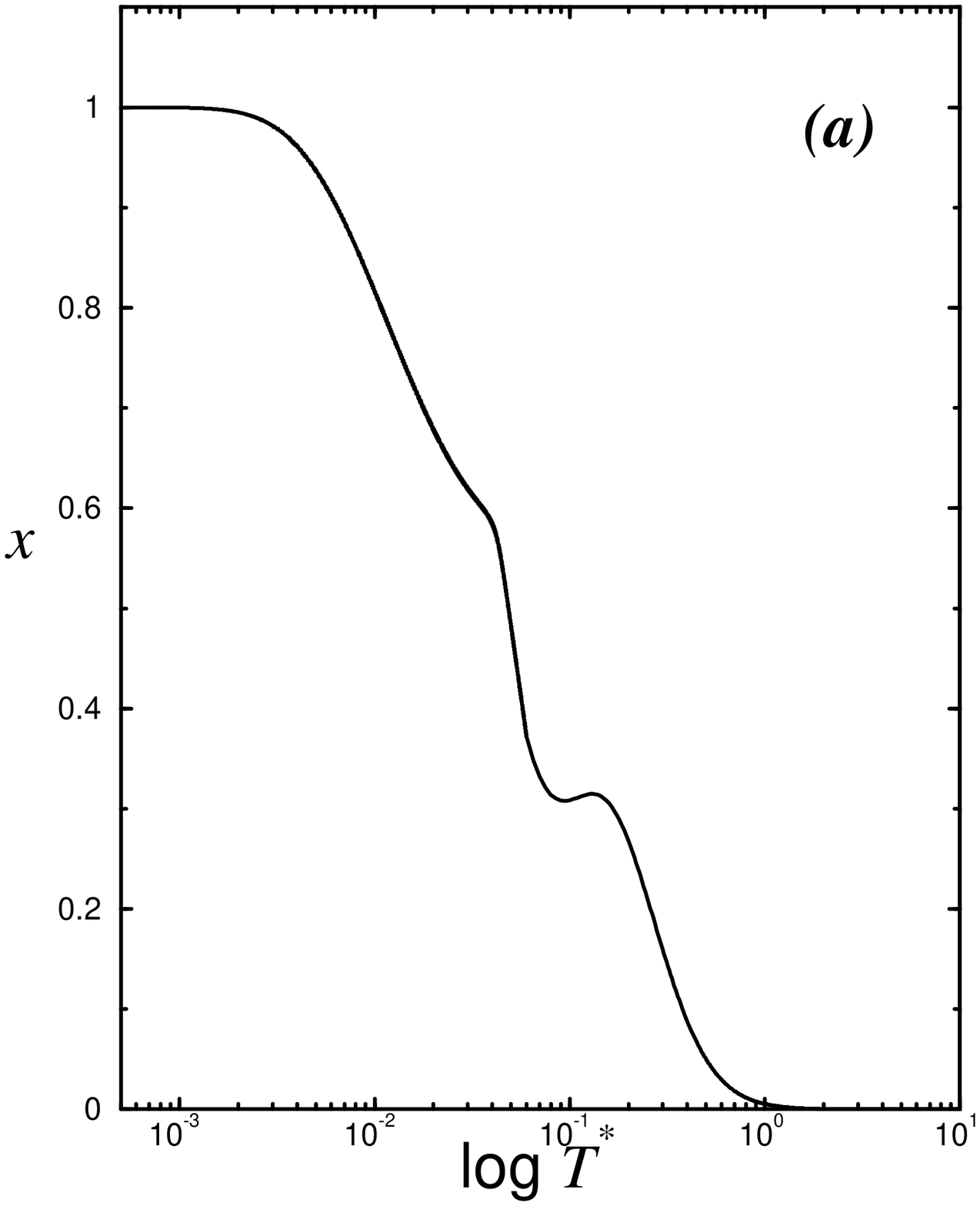}
\end{figure}

\begin{figure}[h]
\vspace*{6cm}
\includegraphics{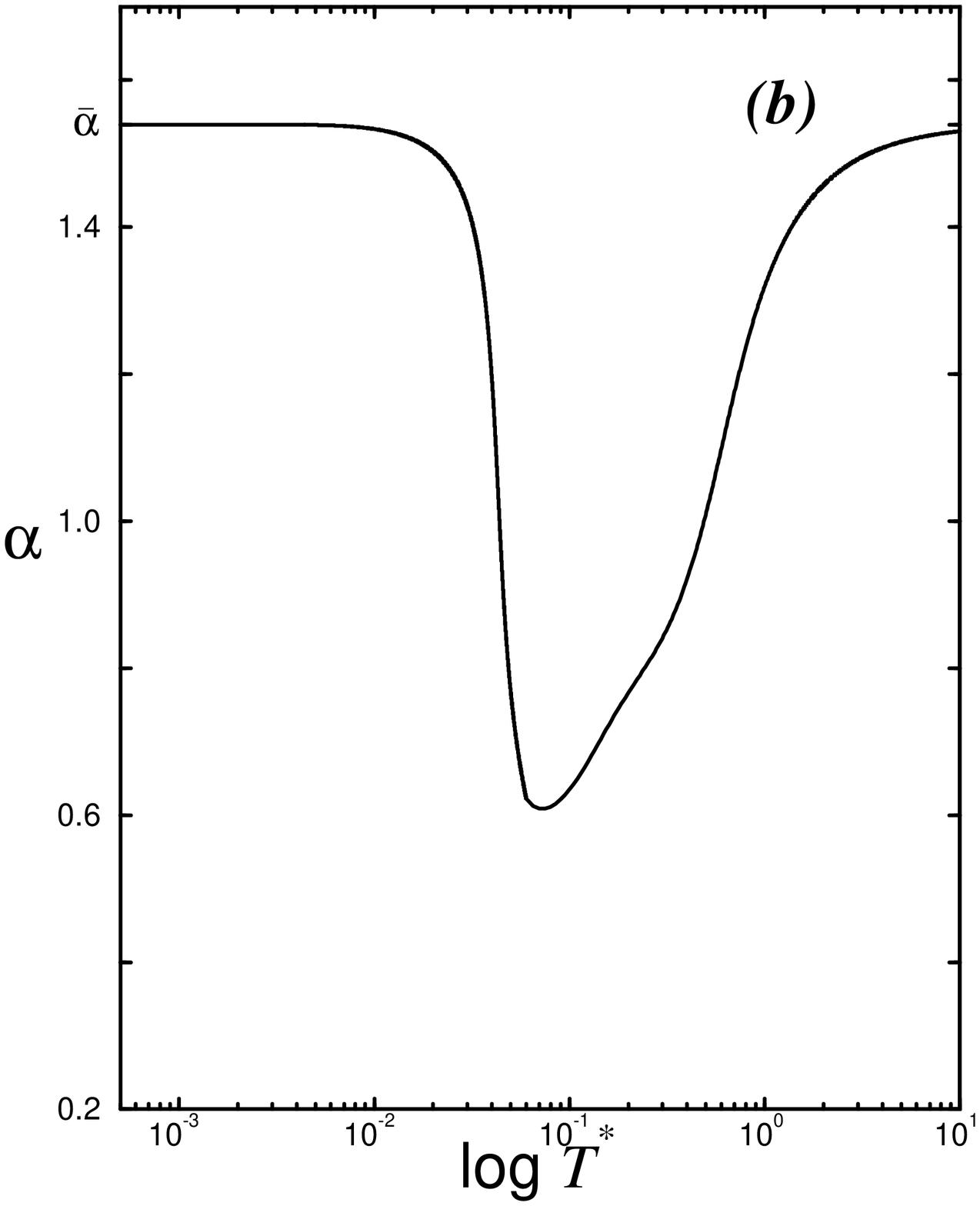}
\end{figure}

For a quench on vapor side, as the temperature is decreased, the density of the salt inside 
$V_i$, $\rho_s$, exhibits an interesting behavior. Even if the density outside $V_i$ is
fixed at the vapor phase, the density inside $V_i$ is at $\rho^l_{\sigma}(T^{\ast})$ 
(see Fig.\ref{VSCC1}(c)). The salt actually separates in two phases, a 
liquid phase inside the globule and a vapor phase outside. Therefore, even 
for a very low density of the salt outside the globule, $\rho_0=\rho^v_{\sigma}(T^{\ast})$, 
the density of salt inside is quite large ($\rho_s=\rho^l_{\sigma} (T^{\ast})$).

Finally, the density of the salt was fixed at $\rho_{c}^{\ast} < \rho_0^{\ast}=0.00698 < \bar \rho$ 
(at $T^{\ast}=0.0618$), which implies a quench on the liquid side of the coexistence 
curve, but in the region where the collapse transition is absent. In this 
case, like in the previous situation, the polymer is in the dilute globular state 
for almost all temperatures. Differently from what has happened on  the vapor side of the 
coexistence curve, at 
low temperatures there is a smooth crossover from the globular state to the
extended state.

As we pointed out in the last two cases, when the density of the salt is small, $\rho_0 < \bar \rho$, the
$DH$ contribution is governed by the density $\rho$ of polymer. For large temperatures, 
there are almost no dipoles, $x\rightarrow 0$ and the density of salt inside the 
volume $V_i$ approaches the value outside $\rho_0$. In this region $z$ 
becomes small and the DH limiting law can be used in Eq.(\ref{press}), which becomes
\begin{equation} 
\label{press0} 
-\frac{3(\alpha^2 -1)}{4\pi a^3 \alpha^3 N^{3/2}} +\frac{W_1}{2}\rho^2 + W_E \rho^{3/2} =0\;,
\end{equation}
where $W_E \equiv fa^{3/2}\sqrt{4 f\pi}/6 (T^{\ast})^{3/2} $. In the 
$N\rightarrow \infty$ limit, the solution of Eq.(\ref{press0}) implies a 
finite value for the density $\rho$ and, consequently, the size of the polymer scales as 
\begin{equation} 
\label{scalea0}
R=a_0 N^{1/3}\;,  
\end{equation}
where $a_0 \equiv 3T^{\ast}\left(W_1^{\ast}/4\pi \right)^{2/3} a/f(1-x)$,
with $W_1^{\ast}\equiv W_1/a^3$ as a dimensionless parameter, corresponding 
to the quality of the solvent. 

Bellow $T_{\zeta}^{\ast}$ the conformation of the $PA$ is governed by 
branch $l$ (see Appendix \ref{Energy}). Since along this branch all the monomers are neutralized by the 
counterion, the attractive electrostatic interaction
is absent and in Eq.(\ref{press}) the hard-core repulsive term
(repulsive) has to be balanced by the elastic term (attractive). In this case, 
we find that
\begin{equation} 
\label{scale2}
-\frac{3(\alpha ^2-1)}{4\pi a^3 \alpha ^3N^{3/2}}+\frac{W_1}2\rho ^2+W_2\rho ^3 = 0\;,
\end{equation}
which is the same as for a neutral polymer: 

\begin{itemize} 
\item[(i)]  If $W_1\gg 0$ (good solvent), the elastic term $-3(\alpha^2 -1)/4\pi a^3
\alpha^3 N^{3/2}$ balances the hard-core contribution $W_1\rho ^2/2$ and gives 
a radius of Gyration $R=a_1N^{3/5}$, where the amplitude $a_1\equiv \left( 3 W_1^{\ast}/8\pi 
\right) ^{1/5}a$. So, the $PA$ is in an extended state;
\item[(ii)]  if $W_1=0$ (theta solvent), the second virial term
is zero, and the elastic term balances the third virial $W_2\rho ^3$. In this case the radius
scales as $R=a_2N^{1/2}$, where $a_2\equiv \left( 9
W_2^{\ast}/(4\pi )^2 \right)^{1/8}a$, and $W_2^{\ast}\equiv
W_2/a^6$. In this
case, the $PA$ behaves like a coil;

\begin{figure}[h]
\vspace*{6cm}
\includegraphics{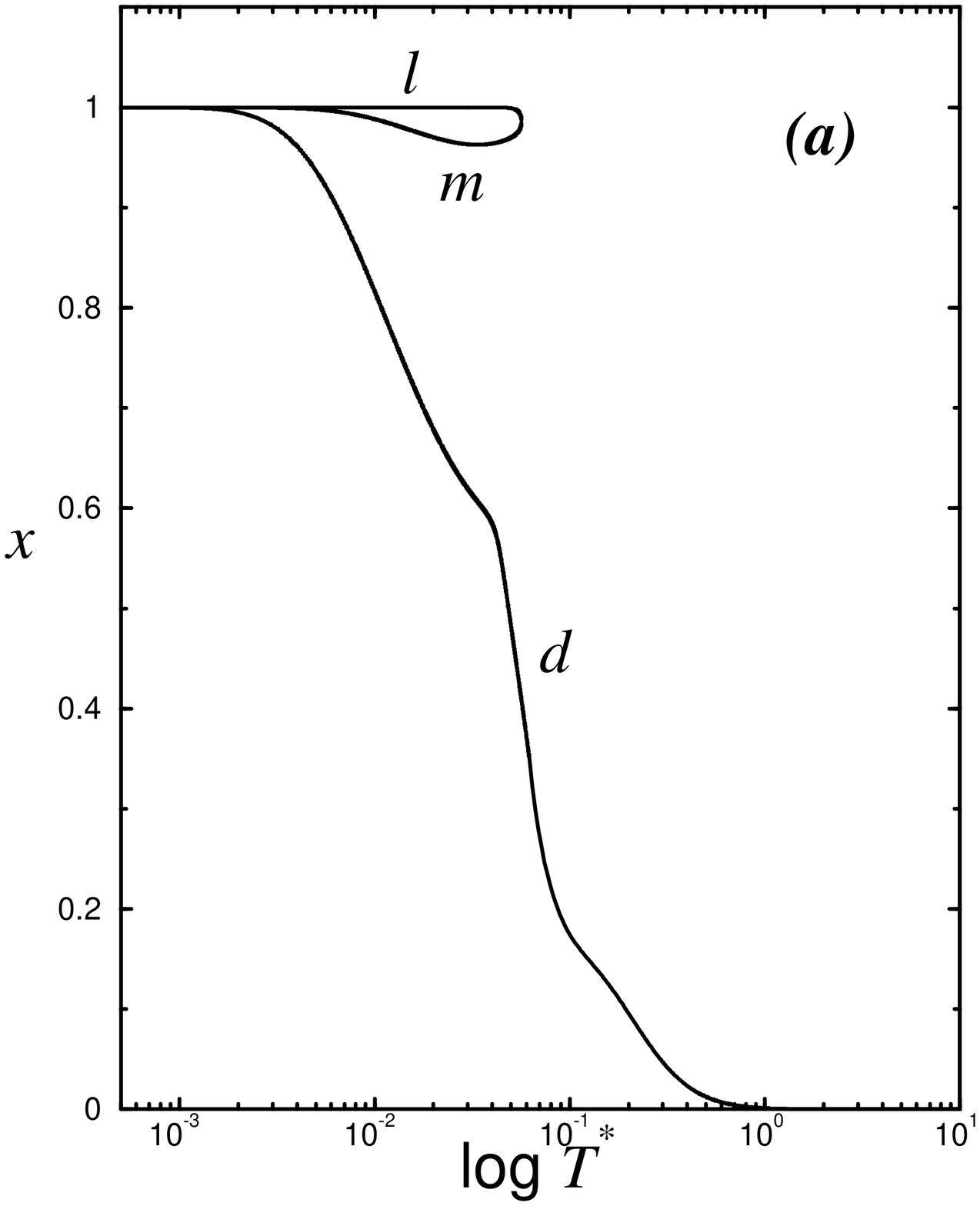}
\end{figure}

\begin{figure}[h]
\vspace*{1cm}
\includegraphics{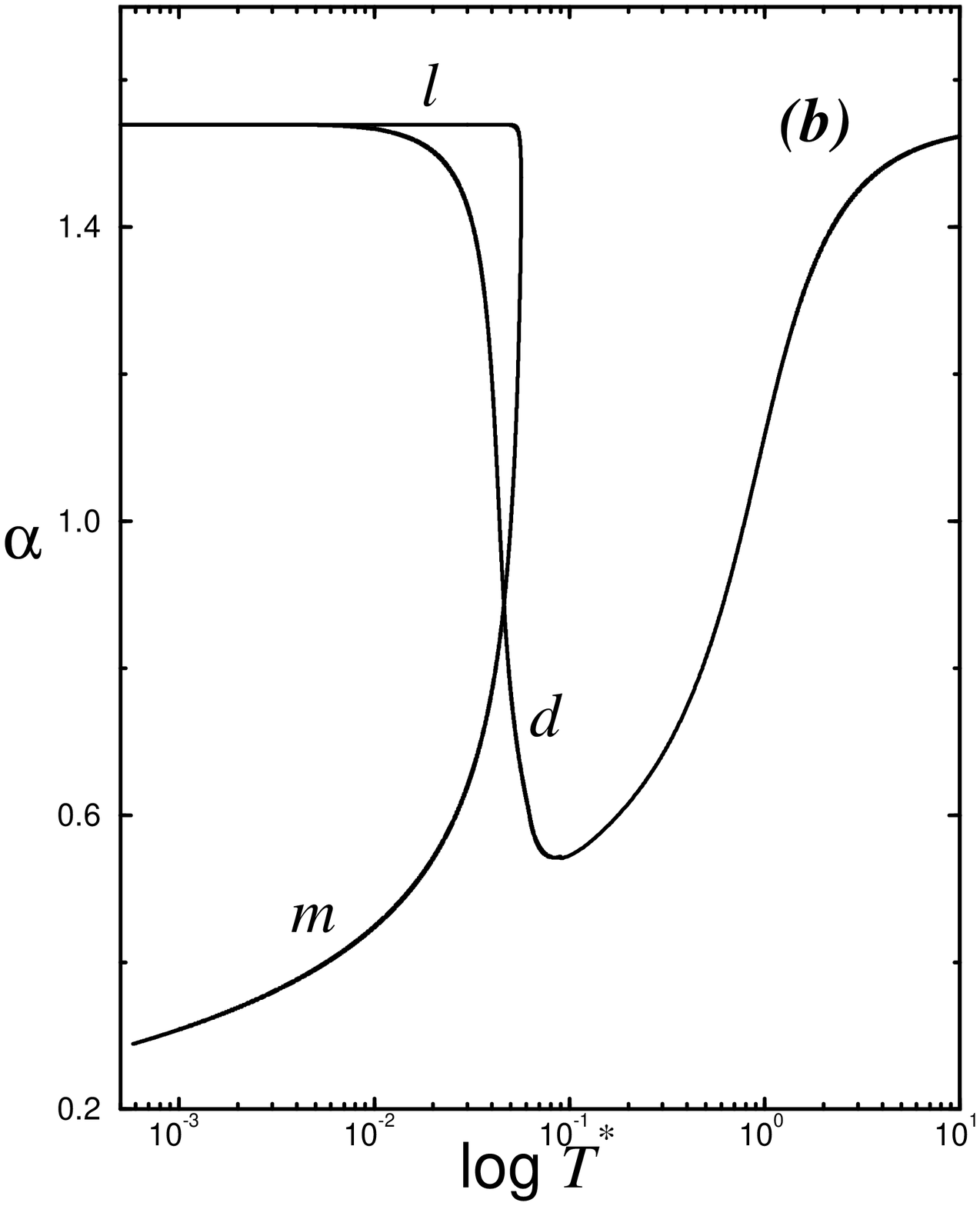}
\end{figure}

\begin{figure}[h]
\vspace*{10.3cm}
\includegraphics{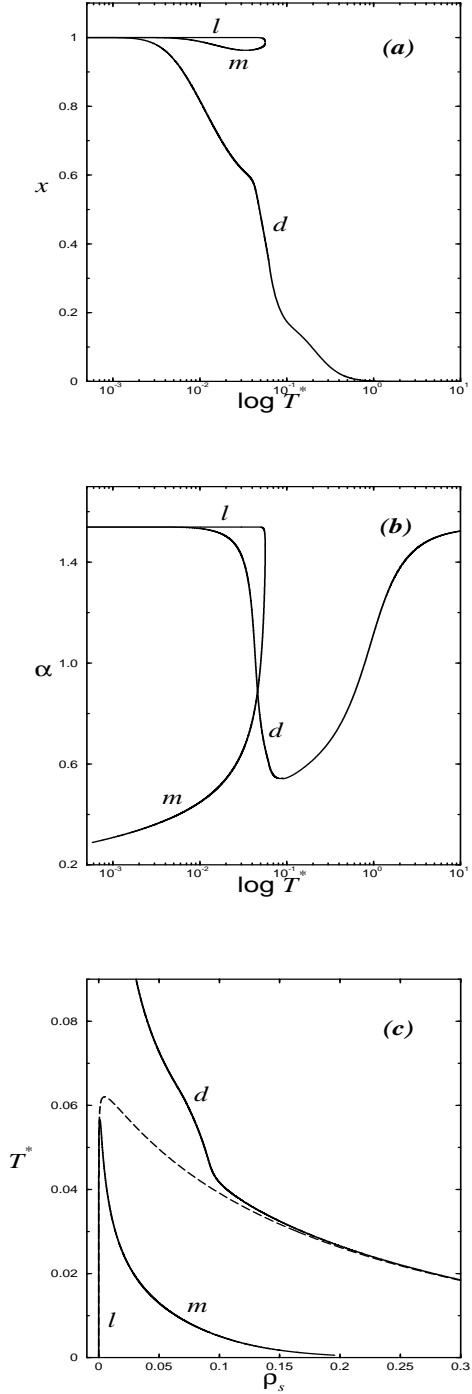}
\caption{(a) Fraction of dipoles $x$, (b) expansion factor $\alpha$ and 
(c) density of salt inside of spherical region, $\rho_s$, for a quench on 
the vapor side of the coexistence curve, is shown against the temperature, for 
$N=100$, $f=1$ and $W_1 =4\pi a^3 /3$. The stable branch $d$ represents a dilute collapsed state,
$l$ represents a stable extended state and $m$ is a metastable state. The dashed 
line in (c) represents the coexistence boundary. Notice how the density 
of salt inside the globule asymptotes to the liquid side of the phase boundary.}
\label{VSCC1}
\end{figure}

\item[(iii)]  if $W_1\ll 0$ (poor solvent), the second virial
coefficient $|W_1|\rho ^2/2$ balances the third virial $W_2\rho^3$ and 
the radius of Gyration scales as $R=a_3N^{1/3}$, where $a_3\equiv
\left(3 W_2^{\ast}/2\pi |W_1^{\ast}|\right)^{1/3}a$. Thus, the $PA$
is in a dense collapsed state. 
\end{itemize}
Unfortunately this interesting behavior will be impossible to observe experimentally 
since $T_{\zeta}^{\ast}$ is too low.

\bigskip
\section{\bf CONCLUSION}
\label{conc}
\bigskip

In this work, we have analyzed the behavior of a neutral polyampholyte in the 
presence of salt, using the full Debye-H$\ddot u $ckel-Bjerrum-Flory ($DHBj$F) theory. 
Allowing for the association between the charged monomers and the free ions of salt, we found that when 
the density of salt is large $ \rho_0 > \bar \rho$, at high 
temperatures the $PA$ is  extended. As the temperature is decreased, 
there is a collapsed transition to a dilute collapsed state at $T=T_{\theta}$. 
As the temperature is decreased even further, the free ions of the
salt associate with the charged monomers, decreasing
the electrostatic attraction. In fact, at zero temperature, 
all of  the monomers are neutralized and, since there is no free charges
left, the chain  behaves as a usual neutral polymer. Therefore, in the
presence of a good solvent, there is a crossover from a
dilute globular state to an extended state at low temperatures.
If the density of salt is small, $ \rho_0 < \bar \rho$, the polymer is
in the dilute collapsed state for almost all temperatures. A
first-order phase transition to a low temperature extended state occurs
for an unphysical low temperature.

\acknowledgments

This work was supported in part by CNPq - Conselho Nacional de
Desenvolvimento Cient\'{\i}fico e Tecnol\'ogico and FINEP
Financiadora de Estudos e Projetos, Brazil.

\appendix
\section{\bf Coexistence Curve for the salt}
\label{CC}

The salt solution exhibits two possible phases, liquid (high density) and 
vapor (low density). The coexistence curve boundary $\rho_{\sigma}(T^{\ast})$ 
between these phases is given by the solutions of the coupled equations
\begin{equation} 
\label{CC1} 
P(\rho_{\sigma}^v )=P(\rho_{\sigma}^l)\;,
\end{equation}
and
\begin{equation} 
\label{CC2} 
\mu (\rho_{\sigma}^v )=\mu (\rho_{\sigma}^l )\;,
\end{equation}

where $P \equiv -\partial \beta F /\partial V$ and $\mu \equiv
\partial \beta F / \partial N$. The Helmoltz free energy
$\beta F = \beta F_{HC} + \beta F_{DH} +\beta F_{id}$
is given by the Eqs.(\ref{Fhc0}), (\ref{FDH}) and (\ref{MIX0}). Explicitly,
\begin{equation} 
\label{Pexp}
P (\rho_{\sigma})= \rho_{\sigma} +\frac{B\rho_{\sigma}^2}{1-B\rho_{\sigma}} + \frac{1}{8\pi a^3}
\biggl[2 \ln (z +1) - \frac{z (z +2)}{z +1}\biggr] 
\end{equation}
and
\begin{eqnarray} 
\label{muexp}
\mu (\rho_{\sigma} )=&-&\frac{1}{2T^{\ast}}\frac{z}{z +1} + \ln \Lambda^3 \frac{\rho_{\sigma}}
{2}- \ln (1-B\rho_{\sigma})\nonumber\\
& +&\frac{B\rho_{\sigma}}{1-B\rho_{\sigma}}\;, 
\end{eqnarray}
where $z=\sqrt{4\pi \rho_{\sigma}^{\ast} /T^{\ast}}$. The resulting coexistence curve, for 
$B=4a^3 /3\sqrt{3}$ (bcc packing), is shown in Fig.\ref{CCurve} and, for low 
temperatures, the vapor and liquid branches are given analytically by
\begin{eqnarray}
\label{rhov}
\rho_{\sigma}^v \approx &&\exp \biggl[ -\frac{1}{2T^{\ast}} +
\frac{1}{2}\sqrt{\frac{1}{\pi \rho_M T^{\ast}}} 
+\left( -\frac{1}{2}+\frac{1}{8\pi \rho_M}\right) \ln T^{\ast}\nonumber\\
&&\qquad \quad -\ln 4\sqrt{\frac{\pi}{\rho_M}}-\frac{\ln 4\pi \rho_M}{8\pi \rho_M}\biggr]
\end{eqnarray}
and 
\begin{eqnarray}
\label{rhol}
\rho_{\sigma}^l \approx &&\rho_M \biggl[ 1-4\sqrt{\pi \rho_M T^{\ast}} -2T^{\ast}\ln
\frac{1}{T^{\ast}}\nonumber\\
&&\qquad \quad -\left( 2\ln 4\pi \rho_M -2-8\pi \rho_M \right) T^{\ast}\biggr]\;,
\end{eqnarray}
where $\rho_{M}\equiv 1/B$ is the maximum  packing density. As the temperature 
$T^{\ast}\rightarrow T^{\ast}_c$,  $\rho_{\sigma}^l - \rho_{\sigma}^v \to 0$ and at the critical point, 
$\rho_{\sigma}^v =\rho_{\sigma}^l =\rho_c^{\ast}$. The critical temperature and density are 
$T_c^{\ast} \approx 0.0624$ and $\rho_c^{\ast} \approx 0.00462$. In this case, the reduced
temperature is only 10$\%$ higher than the MC estimate 
($T^{\ast}_c =0.057 \pm 1_5,\; \rho^{\ast}_c =0.030 \mp 8$)
\cite{Panan92}, but the critical density is too small. The low value of 
critical density is due to the fact that $DH$ theory omits important
nonlinear configurations, such as formation of cluster, dipoles, quadrupoles, etc.
This configurations can be reintroduced into the theory by allowing for the existence
of equilibrium between monopoles and clusters.
At the simplest level that we are considering  in this paper, salt dipoles
are taken as neutral species that do not interact electrostatically with 
$PA$, and can be ignored.

\section{\bf The Temperature $T_{\theta}^{\ast}$}
\label{Temperature}
 
In this Appendix we study the collapsed transition present at high 
temperatures and high concentration of salt. We also compare our 
results with those of Higgs and Joanny\cite{Hi91}. At high temperatures, 
the density of salt $\rho_s$ inside the volume $V_i$ is slightly higher than 
$\rho_0$ (see Fig.\ref{LSCC1}) and, consequently, we can write
\begin{equation}
\label{rhos}
\rho_s =(1 + \delta )\rho_0\;,
\end{equation}
where $\delta \ll 1$. 

The density of free charges along the polymer can always be defined as 
$\rho =\gamma \rho_0$. In this case, the total density of charges inside $V_i$ is 
\begin{equation}
\rho_1 = \rho_0 (1+\delta + \tilde \gamma )\;,
\end{equation}
where $\tilde \gamma = f\gamma$. 

If $\rho_0 < \bar \rho$, $\tilde \gamma > 1$ and, therefore, at high temperatures, where association becomes 
irrelevant, the electrostatic contribution to the free energy scales as $\rho^{3/2}$. This attractive interaction 
is balanced by second virial term $W_1 \rho^2 /2$, which forces $PA$ to collapse.

For high concentration of salt, $\tilde \gamma < 1$. Therefore, in this case, $\delta$ and 
$\tilde \gamma$ can be treated as small parameters. Solving 
Eq.(\ref{chem}) for $\delta$ up to second order in $\tilde \gamma$ gives 
\begin{eqnarray}
\label{delta}
\delta =&& \tilde \gamma \;\frac{1}{4 T^{\ast}}\frac{z_0}{(z_0 +1)^2}
\biggl[\frac{1}{(1-B\rho_0 )^2}-\frac{1}{4 T^{\ast}}\frac{z_0}
{(z_0 +1)^2}\biggr]^{-1}\nonumber\\
&&-{\tilde \gamma}^2 \;\frac{1}{16 T^{\ast}}\frac{z_0}{(z_0 +1)^4 (1-B\rho_0)^3}
\biggl[ \frac{1}{(1-B\rho_0 )^2}\nonumber\\
&&\qquad \quad -\frac{1}{4 T^{\ast}}\frac{z_0}{(z_0 +1)^2}\biggr]^3 \biggl[ \frac{(3z_0 +1)(z_0 +1)}{1-B\rho_0}
\nonumber\\
&&\qquad \quad -\frac{1}{2T^{\ast}}z_0 (1-3B\rho_0)\biggr]\;.
\end{eqnarray}
Assuming that both the monomers and the ions of salt are point particles 
($B=0$) and expanding in $z_0$, to leading order the expression above reduces to Eq.(3.7) of Higgs and Joanny's paper\cite{Hi91}. 
Unlike their analysis, however, we find that when the full $DH$ theory is used, the unphysical divergence 
at $1/(1-B\rho_0)^2 = z_0/4T^{\ast}(z_0+1)^2$ is actually located inside the coexistence curve. Since the density of salt is always outside the coexistence 
region, this divergence is never encountered. 

The scaling form of the radius of gyration is then given by substituting Eq.(\ref{delta}) into 
Eq.(\ref{press})
\begin{eqnarray}
&-&\frac{3(\alpha ^2-1)}{4\pi a^3 \alpha ^3N^{3/2}}+ \frac{W_1}2\rho ^2 + W_2\rho ^3\nonumber\\ 
&-&\frac{\pi f^2}{2(T^{\ast})^2} \frac{a^3}{z_0 (z_0 +1)^2 (1-B\rho_0)^2}\nonumber\\
&&\times \biggl[ \frac{1}{(1-B\rho_0 )^2}-\frac{1}{4 T^{\ast}}\frac{z_0}{(z_0 +1)^2}\biggr]^{-1}\rho^2 = 0\;.
\end{eqnarray}
Note that this expression is just the minimization of the free energy of a usual polymer with 
the second virial coefficient renormalized to
\begin{eqnarray}
{\tilde W_1} \equiv &&\frac{W_1}{2} - \frac{\pi f^2}{2(T^{\ast})^2} \frac{a^3}{z_0 (z_0 +1)^2 
(1-B\rho_0)^2}\nonumber\\
&&\qquad \quad \times \biggl[ \frac{1}{(1-B\rho_0 )^2}-\frac{1}{4 T^{\ast}}\frac{z_0}{(z_0 +1)^2}\biggr]^{-1}\;.
\end{eqnarray}
As in conventional polymers, for $\tilde W_1 >0$ (good solvent case), the polymer will be fully 
extended, while for $\tilde W_1 <0$ (bad solvent case) will assume a globular form. At 
$T^{\ast}=T_{\theta}^{\ast}$, given by $\tilde W_1 =0$ in equation above (theta solvent), 
there is a collapse transition, as shown in Fig.\ref{LSCC1}.

\section{\bf The first-order phase transition at low temperatures when $\rho_0<\rho_{c}$} 
\label{Energy}

If the density of salt is chosen to be on the vapor side of the 
coexistence curve, at very low temperatures Eqs.(\ref{press}), (\ref{chem}) 
and (\ref{reac}) exhibit two possible solutions: one with $\rho_s=\rho^v$ 
(low density of salt) and another with $\rho_s=\rho^l$ (high density of salt) 
as $T^{\ast} \rightarrow 0$. The first solution exists only for low temperatures, 
while the second is present for all temperatures. 
Consequently, we have to investigate if there would be a phase transition 
between these two phases as the temperature is decreased.
In order to address this question, we have to compare the respective free energies for the two branches.

The free energy associated with the solution with low density of salt is 
obtained by substituting 
\begin{eqnarray} 
\label{sol1} 
x&\approx& 1 + {\cal O}(e^{-1/2T^{\ast}})\;,\nonumber\\
\alpha&\approx&\bar{\alpha} + {\cal O}(e^{-1/2T^{\ast}})\;,\nonumber\\
\rho_s&\approx& \rho^v + {\cal O}(e^{-1/2T^{\ast}})\;,
\end{eqnarray} 
into the Eq.(\ref{gibbs}) what leads to
\begin{equation} 
\label{freextend}
\beta {\cal G}^E \approx - \frac{N f}{2 T^{\ast}}
-\frac{N f}{2\sqrt{\pi \rho_M T^{\ast}}}+ {\cal O}
\left(e^{-1/2T^{\ast}}\right)\;.
\end{equation} 

On the other hand, the free energy associated with the high density state is obtained by substituting 
\begin{eqnarray} 
\label{sol2}
x&\approx& 1+{\cal O}\left(\frac{e^{-3/4\sqrt{\pi \rho_M T^{\ast}}-\ln T^{\ast}/8\pi \rho_M}}
{\sqrt{T^{\ast}}}\right)\;,\nonumber\\ 
\alpha&\approx&\bar{\alpha}+{\cal O}(1-x)\;,\nonumber\\ 
\rho_s&\approx& \rho^v+{\cal O}(1-x)\;, 
\end{eqnarray} 
into Eq.(\ref{gibbs}). The resulting free energy is given by
\begin{eqnarray} 
\label{freecollap}
\beta {\cal G}^C \approx &&- \frac{N f x}{2 T^{\ast}}
-\frac{N f}{2\sqrt{\pi \rho_M T^{\ast}}}\nonumber\\
&&\qquad \quad + {\cal O}\left(\frac{e^{-3/4\sqrt{\pi \rho_M T^{\ast}}-\ln T^{\ast}/8\pi \rho_M}}
{\sqrt{T^{\ast}}}\right)\;.
\end{eqnarray} 

Comparing Eqs.(\ref{freextend}) and (\ref{freecollap}) gives 
that $\beta {\cal G}^C \approx \beta {\cal G}^E + (1-x)f N/2T^{\ast}$,
where $(1-x)$ scales as $\exp\{-3/4\sqrt{\pi \rho_M T^{\ast}}-\ln T^{\ast}/8\pi \rho_M\}/\sqrt{T^{\ast}}$. Since the 
last term in the above expression is positive for low temperatures, the 
phase of low density of salt is more stable at low temperatures. This 
indicates that there is a phase transition between the
dilute globular state and the extended state. Unfortunately,
numerical analysis indicates that this transition happens at very low temperature.


\begin{references}
\bibitem{KEbel} For $T^{\ast}>0.2$ we have used the modified equilibrium constant derived by Ebeling 
$$
K^{Eb}(T^{\ast})=\frac{2\pi a^{3}}{3(T^{\ast})^3}\biggl[Ei\left(\frac{1}{T^{\ast}}\right)-Ei\left(-\frac{1}{T^{\ast}}\right)+6T^{\ast}+4(T^{\ast})^3\biggr] $$
$$-\frac{2\pi a^3}{3}e^{1/T^{\ast}}\biggl[2+\frac{1}{T^{\ast}}+\frac{1}{(T^{\ast})^2}\biggr]$$
$$-\frac{2\pi a^3}{3}e^{-1/T^{\ast}}\biggl[2-\frac{1}{T^{\ast}}+\frac{1}{(T^{\ast})^2}\biggr]\;.
$$
\end{references}
\end{document}